\documentclass{aa}  

\usepackage{graphicx}
\usepackage[]{natbib}
\usepackage{txfonts}
\usepackage[utf8]{inputenc}
\newcommand{\feh}{$\rm [Fe/H]$}

\newcommand{\aabun}{${\rm [\alpha/Fe]}$}
\newcommand{\mgfeabun}{${\rm [Mg/Fe]}$}
\newcommand{\sifeabun}{${\rm [Si/Fe]}$}
\newcommand{\cafeabun}{${\rm [Ca/Fe]}$}
\newcommand{\tifeabun}{${\rm [Ti/Fe]}$}

\newcommand{\sicafeabun}{${\rm [SiCa/Fe]}$}
\newcommand{\mgabun}{${\rm [Mg/H]}$}

\begin{document} 

 \title{From globular clusters to the disc: the dual life of our Galaxy}
%\subtitle{ Bulge low-$\alpha$ element stars}
\author{Alejandra Recio-Blanco\inst{1}          }
\institute{
         Universit\'e C\^ote d'Azur, Observatoire de la C\^ote d'Azur, CNRS, Laboratoire Lagrange, France \\  \email{arecio@oca.eu}         }
         
   \date{Received ...; accepted ...}
   \abstract {The halo and disc globular cluster population can be used as a tracer of the primordial epochs
   of the Milky Way formation. In this work, literature data of globular clusters
ages, chemical abundances, and structural parameters are studied,
explicitly focussing on the origin of the known split in the age-metallicity relation (AMR) of globular clusters.
When the $\alpha$-element abundances, which are less strongly affected by the internal light-element spread of
globular clusters (Si, Ca), are considered, a very low observational scatter among metal-poor clusters is observed. A plateau at \sicafeabun$\sim$0.35~dex, with a dispersion of only 0.05~dex (including abundance errors) is observed up to a metallicity of about
-0.75~dex. Only a few metal-poor clusters in this metallicity interval present low \sicafeabun~ abundances. 
Moreover, metal-rich globular clusters show a knee in the \aabun~ versus \feh~ plane around \feh$\sim$-0.75~dex. 
As a consequence, if a substantial fraction of galactic globular clusters has an external origin, they have to be mainly formed either 
in galaxies that are massive enough to ensure high levels of $\alpha$-element abundances even at intermediate metallicity, or 
in lower mass dwarf galaxies accreted by the Milky Way in their early phases of formation. 
%The {\it in situ} formation of those clusters with high \aabun~ values is also compatible with the results. 
Finally, clusters in the metal-poor branch  of the AMR present an anti-correlation of \sicafeabun~ with the 
total cluster magnitude, while this is not the case for metal-rich branch clusters. In addition, this lack of
faint high-$\alpha$ clusters in the young metal-poor population is in contrast with what is observed for old
and more metal-poor clusters, possibly reflecting a higher heterogeneity of formation environments at lower metallicity.
Accretion of high-mass satellites, as a major contribution to the current Milky Way globular cluster system both in the metal-poor and the metal-intermediate 
regime is compatible with the observations.
}

\maketitle
   
%
%-------------------------------------------------------------------

\section{Introduction}
In the era of Milky Way surveys, photometric, spectroscopic, and astrometric data for large
numbers of stars are rapidly changing our view of the Galaxy. The fingerprints of 
different galactic evolutionary processes can now be revealed in the chemodynamical
characteristics of the stellar populations, with databases increasing every day.
The isolated, time-frozen snapshots of separated galactic components today
become a moving picture that progressively unveils the Milky Way history.
In this context, the comparative analysis of what looked like, in principle, different populations
(e.g. the halo and the disc) is crucial to detect the signatures of major evolutionary events that
globally affect  the Galaxy formation.

Globular clusters (GC) have always been a key population for Galactic archaeology studies.
They are witnesses of the ancient times of the Milky Way history, with ages
spanning over five gigayears and overlapping the halo and the primordial disc formation.
The galactic globular cluster system  is therefore at the cross-roads of two key Milky Way
stellar populations, and it is precisely this characteristic that this analysis wishes to
highlight.

Moreover, GC formation points out the fundamental question of galaxy {\it \textup{in situ}} versus
{\it \textup{ex situ,}} or accretion formation. In simulations, the  {\it \textup{in situ}} and/or {\it \textup{ex situ}} origin of
a population can be tagged and tracked separately. However, the situation is
much more complex in observations. The distinction between GCs  that formed {\it \textup{in situ}}  and those
formed in another galaxy that were subsequently accreted is blurred when detailed physical
processes are taken into consideration. For instance, clumpy dissipative collapse, inflowing of
cold gas into a turbulent gas disc, mergers of gas-rich galaxies originating {\it \textup{in situ}} formation,
etc. are processes that highlight the weakness of a sharp classification {\it \textup{into in situ}} versus accretion scenarii.

%The goal of this work is to examine the duality of the Milky Way globular clusters population
%in the light of the different accretion processes invoked to explain the galactic disc characteristics.

In recent years, several studies have analysed the properties of the Galactic GC age-metallicity
relation (AMR). First of all, the increase in age dispersion as a function of the metallicity has been noted
\citep[e.g.][]{Buonanno98, Alfred99, VandenBerg00, SalarisWeiss02, Francesca}. Later on,
\cite{Antonio09}, \cite{Forbes10}, and \cite{Dotter11} reported that the AMR exhibited a split, with one sequence
of old clusters at all metallicities, and another sequence of intermediate-metallicity clusters.
GC accretion has currently been invoked to explain the spread of the AMR. 

More recently, \cite{Leaman13} used the age estimates of \cite{V13} to reveal that the AMR bifurcates at
\feh$\sim$-1.8 dex, with clusters having halo type or disc-like orbits that populate different branches.
\cite{Leaman13} suggested that the metal-rich branch of the AMR consists of clusters
that formed in situ in the disc, while the metal-poor clusters could have been formed in relatively low-mass
(dwarf) galaxies and later accreted by the Milky Way.
More generally, it is currently assumed that metal-rich globular clusters are mainly an
in situ\textup{} population because the stellar components of massive early-type galaxies and their
red GCs share many physical properties \citep[e.g.][]{Pota13}. In this sense, metal-poor
GCs seem to be better candidates for accretion. 

Nevertheless, the real observational situation is much more complex. In particular, radial
chemical gradients have been detected in several GC systems \citep[e.g.][]{HarrisGradients},
being probably similar for both metal-rich and metal-poor GCs \citep[][]{Forbes11}. Although these gradients
appear to be related with some sort of {\it \textup{in situ}} formation, the invoked scenario should be able to explain them, regardless of which scenario it is. In parallel, many clear signatures of accretion
in the halo have been revealed, and the Sagittarius dwarf galaxy and its associated
streams and globular clusters is an outstanding example of it.

From the point of view of galactic disc formation, the question of the {\it \textup{in situ}}  versus accretion 
contribution is a key topic as well. In particular, the thick- or thin-disc bimodality has been offered to be explained by either
accretion processes such as minor mergers \citep[e.g.][]{Villalobos} or gas infall (\cite{Cristina2Infall}, or
by poor internal secular processes such as radial migration \citep[e.g.][]{Schonrich2009}. In this sense, the
apparent membership of many metal-rich globular clusters to the galactic thick disc should once again
be linked to these disc formation mechanisms.

 In this work, I address the question of the {\it \textup{in situ}} versus accretion GC formation by focussing on the 
split of the AMR:  are there signs of accretion in the chemical properties of the clusters, and what is
their link with the AMR? How do other properties such as the cluster total luminosity vary in the
different features of the AMR? What is the transition between the halo and the disc chemical
properties? It is important to note that the complex formation and disruption mechanisms
analysed through theory and simulations \citep[for an extensive review, see][]{Forbes18} hinder
the interpretation of present-day GC properties.% as being shaped mostly by formation or by disruption.

%if, as suggested, this bifurcation separates in situ from accreted globular clusters , 
%can we say something about the accretions nature? Can this be linked to ancient
%accretion processes currently invoked to explain the Milky Way disc and,
%in particular, the thick$-$thin disc duality? Are there other signs of accretion in
%the clusters chemical properties? Why do we observe a two-branches AMR
%rather than an increasing age spread with metallicity?

 With these caveats in mind, literature data of globular clusters
ages, 
%kinematics, dynamics, 
chemical abundances, and structural parameters are studied here
in the perspective of the bifurcated age-metallicity relation.
In Sect. 2 the observational data are presented. Sect. 3 compares the GC age-metallicity relation with that of local disc stars. Sect. 4 explores
the $\alpha-$element abundances of GCs in relation with the AMR
properties. Sect. 5 presents an analysis of the absolute magnitude distribution of the clusters. %Sect. 6
%includes the analysis of the clusters orbital properties and Sect. 7 focusses on the globular
%cluster kinematics. 
The conclusions are presented in Section~6.
%\vspace{-0.9 cm}

%--------------------------------------------------------------------
\section{Observational data of GCs and disc stars}

This work is based on literature data for the GC
analysis and on data from the AMBRE Project \citep[e.g.][]{Ambre} for the galactic disc comparison sample.

\smallskip
Table 1 presents the complete list of parameters 
and chemical abundances for the GCs considered in this work. In particular, the following compilations have been used:
\begin{itemize}
\item Age estimates have been taken from the homogeneous sample of \cite{V13}, completed by \cite{Leaman13}
using the same procedure.
\item The \feh ~abundance corresponds to the \cite{EugenioScale} scale.
%\item The space velocities and the orbital parameters are those of \cite{Dinescu99a}, \cite{Dinescu99b} and \cite{Dinescu07}.
\item The total luminosity has been taken from the most recent version of the Harris catalogue \citep[][]{Harris}.%, and 
%for the extragalactic clusters from the work of \cite{vdB06}, who uses the compilations of \cite{Sharina05}
%(based on Hubble Space Telescope photometry) and \cite{VdBMakey04}.
\item The mean abundance of $\alpha-$elements with respect to iron, in particular \mgfeabun, \sifeabun,  \cafeabun ,~and  \tifeabun,
have been taken where possible from \cite{EugenioUVES} and other papers of the same group (50\% of the sample), and otherwise, from the most recent estimate from the
literature (cf. Table~1, col. 7)\footnote{Absolute errors in the abundances for this heterogeneous sample from the literature can be of the order of 0.1~dex}. 
\end{itemize}

For the local disc comparison sample, the stars included in the AMBRE Project data have been considered. In particular, the
AMBRE:HARPS sample of \cite{Harps} was used. The corresponding \feh~ and \mgfeabun~ are those of \cite{SarunasAmbre}\footnote{The errors in the abundances are around 0.06~dex \citep[][Table 4, lower panel]{SarunasAmbre}}, while the ages
are taken from \cite{MichaelAmbre} and were determined using Gaia DR1 parallaxes \citep[][]{LindegrenDR1}.
It is worth noting that possible systematic differences between field stars and GCs could be present in the chemical abundances and the ages because of the different assumptions made in their analysis 
by different groups.
%In particular, the orbital properties can be affected by the inclusion or exclusion of a bar/spiral arms, the presence of the LMC, etc... \citep[e.g.][]{Laporte17}.

\setlength{\tabcolsep}{3pt}%
\begin{table*}
%\caption{List of adopted mean $\alpha-$elements  abundances with respect to iron (column 2), mean Si and Ca abundance with respect to iron (column 3) and
%Mg abundance with respect to iron (column 4) with the corresponding reference for each cluster.}
\caption{Adopted parameters for the GCs: metallicity (Col. 2 in dex), age (Col. 3 in Gyr), $\alpha-$element chemical abundances in dex (mean of \sifeabun~ and \cafeabun~ in Col. 4; mean of \sifeabun,\cafeabun, \tifeabun , and \mgfeabun~ in Col. 5; \mgabun~ in Col. 6), number of observed stars (Col. 7) and  their corresponding reference (Col. 8), and
absolute magnitude (Col. 9).}%local standard of rest velocities (col. 9-11 in km/s), maximum distance from the Galactic plane (col. 12 in kpc) and orbital eccentricity (col. 13). }
\centering
{\scriptsize
%\footnotesize
%\begin{tabular}{l c c c c c c c c c c c c c c c c c}
\begin{tabular}{l c c  c c c c l c}
\hline\hline
Cluster    & \feh   & Age   & \sicafeabun&\aabun  &\mgfeabun & nb & Ref. $\alpha-$abundance &    M$_{V}$  \\% &   U     &    V    &    W   &   Zmax   &   e \\
\hline
NGC~104     & -0.76  & 11.75   $\pm$  0.25   &   0.357  &   0.414  &    0.52  & 11 & \cite{EugenioUVES} &  -9.00  \\%   &  75    $\pm$   20 &   -77   $\pm$   16  &   57    $\pm$   14  &   3.1    $\pm$  0.2   &   0.17    $\pm$  0.03 \\
NGC~288     & -1.32  & 11.50  $\pm$   0.38   &   0.391  &   0.379  &     0.45  & 10 & \cite{EugenioUVES}  & -6.74  \\%&  16  $\pm$     09 & -247  $\pm$   18  &   52.9  $\pm$   0.4 &   5.8    $\pm$  0.3   &   0.74    $\pm$  0.06 \\
NGC~362     & -1.30  & 10.75  $\pm$   0.25   &   0.290  &   0.270  &     0.33  & 92 & \cite{Eu13} &  -8.41  \\%&  21  $\pm$     27 &  -278  $\pm$  22  &   -80   $\pm$   20  &   2.1    $\pm$  1.2   &   0.85    $\pm$  0.05 \\
NGC~1261    & -1.27  & 10.75  $\pm$   0.25   &   0.125  &   0.150  &     0.20  & 3 &\cite{Filler12} &  -7.81  \\%&                 &         &       &           &                   \\
NGC~1851    & -1.18  & 11.00   $\pm$  0.25   &   0.355  &   0.308  &     0.37  & 119 &\cite{Eu11} &  -8.33  \\%&  229  $\pm$    35 &   -217  $\pm$   26  &   -106  $\pm$   31  &   7.6    $\pm$  1.4   &   0.69    $\pm$  0.03 \\
\\%NGC~1904    & -1.68  &                  &          &          &           &     &      &  &  120  $\pm$    30 &   -188  $\pm$  29  &   7     $\pm$   34  &   6.2    $\pm$  2.2   &   0.65    $\pm$  0.08 \\
\\%NGC~2298    & -1.82  &                  &          &          &           &     &      &  &  -62   $\pm$   41 &   -208  $\pm$  21  &   97    $\pm$   44  &   6.7    $\pm$  3.5   &   0.78    $\pm$  0.12 \\
\\%Pal~3   & -1.57  &         &                   &          &           &     &       &  -26  $\pm$   101 &   66    $\pm$  81  &   191   $\pm$   77  &   308.4  $\pm$  54.8  &    0.67   $\pm$   0.11 \\
NGC~2808    & -1.18  & 11.00   $\pm$  0.38   &   0.310  &   0.268  &     0.20  & 12 & \cite{EugenioUVES} &  -9.39  \\%&                              &        &         &       &           \\
NGC~3201    & -1.51  & 11.50   $\pm$  0.38   &   0.298  &   0.249  &     0.34  & 13 & \cite{EugenioUVES} &  -7.46  \\%&                              &        &         &       &            \\
NGC~4147    & -1.78  & 12.25   $\pm$  0.25   &   0.43   &   0.390  &     0.42   & 18 &  \cite{Sandro16}&  -6.16  \\%&  77   $\pm$    64 &   -179  $\pm$   65  &   147   $\pm$   15  &    13.1  $\pm$   1.7  &    0.72   $\pm$   0.10 \\
NGC~4590    & -2.27  & 12.00   $\pm$  0.25   &   0.331  &   0.282  &     0.35  & 13 & \cite{EugenioUVES} &  -7.35  \\%&  188  $\pm$   30 &   34    $\pm$  22  &   5     $\pm$   22  &    9.1   $\pm$   1.3  &    0.48   $\pm$   0.03 \\
NGC~4833    & -1.89  & 12.50    $\pm$ 0.50   &   0.405  &   0.333  &     0.37  & 78 & \cite{Eu14b}  &  -8.16  \\%&         &                 &       &          &          \\
NGC~5024    & -2.06  & 12.25    $\pm$ 0.25   &   0.350  &   0.300  &     0.33  & 16 & \cite{Meszaros15} &  -8.70  \\%&  -38  $\pm$   84 &   38    $\pm$   85  &   -75   $\pm$   16  &    24.2  $\pm$   8.1  &    0.40   $\pm$   0.12 \\
NGC~5053    & -2.30  & 12.25     $\pm$ 0.38   &   0.385  &   0.385  &           &  1 & \cite{Luca15} &  -6.72  \\%&        &                 &       &                 &     \\
\\%NGC~5139    & -1.59  &                  &          &          &           &     &      & &  -64  $\pm$   11 &   -254  $\pm$  09  &   4     $\pm$   10  &    1.0   $\pm$   0.4  &    0.67   $\pm$   0.05 \\
NGC~5272    & -1.50  & 11.75     $\pm$ 0.25   &   0.340  &   0.382  &     0.61  & 33 & \cite{EugenioUVES}  &  -8.93  \\%&  -17  $\pm$    24 &   -116  $\pm$   24  &   -126  $\pm$   05  &    8.7   $\pm$   0.5  &    0.42   $\pm$   0.07 \\
NGC~5286    & -1.70  & 12.50  $\pm$   0.38   &   0.36   &   0.407  &     0.55  & 62 & \cite{Marino15} &  -8.61  \\%&       &                 &       &          &          \\
NGC~5466    & -2.31  & 12.50  $\pm$   0.25   &   0.275  &   0.238  &     0.277 & 3 & \cite{Lamb15} &  -6.96  \\%&  248  $\pm$   65 &   -139  $\pm$  64  &   207   $\pm$  19  &    34.1  $\pm$  14.4  &    0.79   $\pm$   0.03 \\
\\%Pal~5   & -1.47  &         &                   &          &           &     &      & &  10   $\pm$   18 &   -262  $\pm$   34  &   -45   $\pm$   17  &    9.0   $\pm$   1.9  &    0.74   $\pm$   0.18 \\
\\%NGC~5897    & -1.94  &                  &          &          &           &    &   &    &  30   $\pm$    31 &   -301  $\pm$   57  &   118   $\pm$   43  &    5.1   $\pm$   1.1  &    0.64   $\pm$   0.10 \\
NGC~5904    & -1.33  & 11.50   $\pm$  0.25   &   0.340  &   0.316  &     0.41  & 14 &\cite{EugenioUVES} &  -8.81  \\%&  -323 $\pm$    31 &   -140  $\pm$   28  &   -204  $\pm$   28  &    18.3  $\pm$   5.5  &    0.87   $\pm$   0.02 \\
NGC~5927    & -0.29  & 10.75   $\pm$  0.38   &   0.090  &   0.128  &     0.230 & 56 &\cite{Bulbo17} &   -7.80 \\% &                &         &       &                 &     \\
NGC~5986    & -1.63  & 12.25   $\pm$  0.75   &   0.300  &          &     0.284 & 25 & \cite{Johnson17}&   -8.44 \\% &      &                 &       &          &            \\
\\%NGC~6093    & -1.64  &                  &          &          &           &    &  &      &  -12  $\pm$   10 &   -285  $\pm$  37  &   -81   $\pm$   25  &    1.5   $\pm$   0.5  &    0.73   $\pm$   0.17  \\
NGC~6101    & -1.98  & 12.25   $\pm$   0.50   &          &        &  &     &       &   -6.91  \\%&         &       &                           &       &     \\
NGC~6121    & -1.18  & 11.50   $\pm$  0.38   &   0.470  &   0.439  &     0.55  &14  &\cite{EugenioUVES} &   -7.20  \\%&  -57  $\pm$   03 &   -193  $\pm$   22  &   -8    $\pm$   5  &    1.5   $\pm$   0.4  &    0.80   $\pm$   0.03 \\
NGC~6144    & -1.82  & 12.75   $\pm$  0.50   &          &          &   &        & &   -6.75  \\%&  170  $\pm$    09 &   -260  $\pm$   38  &   12    $\pm$   28  &    2.4   $\pm$   0.2  &    0.25   $\pm$   0.15 \\
NGC~6171    & -1.03  & 12.00   $\pm$  0.75   &   0.470  &   0.429  &     0.51  &5 &\cite{EugenioUVES} &   -7.13  \\%&  1    $\pm$    11 &   -67   $\pm$   28  &   -40   $\pm$   24  &    2.1  $\pm$   0.2  &    0.21   $\pm$   0.12 \\
NGC~6205    & -1.58  & 12.00   $\pm$  0.38   &   0.420  &   0.378  &     0.44  & 53 &\cite{EugenioUVES} &   -8.70  \\%&  250  $\pm$   36 &   -85   $\pm$   26  &   -117  $\pm$   18  &    13.2  $\pm$   2.1  &    0.62   $\pm$   0.06 \\
NGC~6218    & -1.33  & 13.00   $\pm$  0.50   &   0.387  &   0.389  &     0.52  & 11 & \cite{EugenioUVES}&   -7.32  \\%&  -50  $\pm$  10 &   -98   $\pm$  17  &   -106  $\pm$   14  &   2.3   $\pm$   0.2  &    0.34   $\pm$  0.05 \\
NGC~6254    & -1.57  & 11.75   $\pm$  0.38   &   0.312  &   0.320  &     0.49  & 14 &\cite{EugenioUVES} &  -7.48  \\%&  -84  $\pm$  09 &   -87   $\pm$  22  &   97    $\pm$   19  &    2.4   $\pm$   0.2  &    0.19   $\pm$   0.05 \\
NGC~6304    & -0.37  & 11.25   $\pm$  0.38   &          &          &           & &  &  -7.32  \\%&         &       &                 &         &            \\
NGC~6341    & -2.35  & 12.75   $\pm$  0.25   &   0.275  &   0.194  &     0.13  & 47 & \cite{Meszaros15}&   -8.20  \\%&  26   $\pm$    23 &   -147  $\pm$   15  &   32    $\pm$   18  &    3.8   $\pm$   0.5  &    0.76   $\pm$   0.03 \\
NGC~6352    & -0.62  & 10.75   $\pm$  0.38   &   0.165  &   0.235  &     0.47  & 9 &\cite{Feltzing09}&   -6.48  \\%&       &                 &       &          &            \\
NGC~6362    & -1.07  & 12.50   $\pm$  0.25   &   0.29   &   0.33   &     0.36  & 2 &\cite{Gratton87} &  -6.94  \\%&  81   $\pm$  13 &   -112  $\pm$   18  &   37    $\pm$   14  &    1.4   $\pm$   0.0  &    0.39   $\pm$   0.04 \\
NGC~6366    & -0.59  & 11.00    $\pm$ 0.50   &   0.275  &   0.290  &     0.29  & 5 &\cite{Johnson16}&   -5.77  \\%&        &                 &       &                &     \\
NGC~6397    & -1.99  & 13.00    $\pm$ 0.25   &   0.309  &   0.312  &     0.46  &13 &\cite{EugenioUVES} &   -6.63 \\% &  34   $\pm$   06 &   -91   $\pm$   12  &   -99   $\pm$   11  &    1.5   $\pm$   0.1  &    0.34   $\pm$   0.02 \\
NGC~6426    & -2.15  & 12.25    $\pm$ 0.25   &   0.37   &   0.345  &     0.44  & 4 &\cite{Hanke16}&   -6.69 \\% &          &                 &       &          &            \\
NGC~6496    & -0.46  & 10.75    $\pm$ 0.38   &          &          &          & &  &  -7.23  \\%&          &       &                 &         &       \\
NGC~6535    & -1.79  & 12.75    $\pm$ 0.50   &   0.365  &   0.348  &    0.478  &30 & \cite{Angie17}&  -4.75  \\%&               &         &       &                &      \\
NGC~6541    & -1.82  & 12.50    $\pm$ 0.50   &   0.436  &   0.377  &    0.35   & 3 & \cite{LeeCarney02}&   -8.37  \\%&         &                 &       &          &           \\
NGC~6584    & -1.50  & 11.75    $\pm$ 0.25   &          &          &        &  &  &   -7.68  \\%&  -78  $\pm$  25 &   -354  $\pm$   51  &   -181  $\pm$   39  &    3.1   $\pm$   2.3  &    0.87   $\pm$   0.05  \\
NGC~6624    & -0.42  & 11.25   $\pm$  0.50   &          &          &       &   &  &   -7.49  \\%&          &       &                 &                &      \\
\\%NGC~6626    & -1.28  &                  &          &          &           &      &   &  &  -32  $\pm$   3 &   -55   $\pm$  21  &   -39   $\pm$   16  &    0.6   $\pm$   0.0  &    0.19   $\pm$   0.06  \\
NGC~6637    & -0.59  & 11.00    $\pm$ 0.38   &          &          &      & &     &   -7.64  \\%&          &       &                           &       &     \\
NGC~6652    & -0.76  & 11.25    $\pm$ 0.25   &          &          &      &   &   &   -6.68  \\%&          &       &                           &       &     \\
NGC~6656    & -1.70  & 12.50    $\pm$ 0.50   &   0.37   &   0.34   &   0.39    & 35 &\cite{Marino11}&   -8.50  \\%&   152 $\pm$    5  & -25  $\pm$    21 &    -118 $\pm$    24 &    1.9   $\pm$   0.1  &    0.53   $\pm$   0.01 \\
NGC~6681    & -1.62  & 12.75    $\pm$ 0.38   &   0.43   &   0.41   &   0.52    & 9 &\cite{Omalley17}&   -7.11 \\% &           &                 &       &                 &     \\
\\%NGC~6712    & -1.01  &                  &          &          &           &    &   &    &   98  $\pm$ 6 & -33  $\pm$    11 &    -127 $\pm$   20 &    0.9   $\pm$   0.2  &    0.75   $\pm$   0.03 \\
NGC~6715    & -1.44  & 11.75   $\pm$  0.50   &   0.34   &   0.285  &   0.28    & 76 &\cite{Eu10} &  -10.01 \\%&       &                 &                 &       &    \\
NGC~6717    & -1.26  & 12.50    $\pm$ 0.50   &          &          &        &   &  &  -5.66  \\%&          &       &                 &               &       \\
NGC~6723    & -1.10  & 12.50    $\pm$ 0.25   &   0.33   &   0.28   &   0.23    &  7 &\cite{AlvaroGC}&  -7.84  \\%&            &                 &                  &       &            \\
NGC~6752    & -1.55  & 12.50    $\pm$ 0.25   &   0.386  &    0.366 &     0.500 & 14 &\cite{EugenioUVES} &  -7.73  \\%&  36   $\pm$    05 &   -23   $\pm$   09  &   24    $\pm$   07  &    1.6   $\pm$   0.1  &    0.08   $\pm$   0.02  \\
NGC~6779    & -2.00  & 12.75   $\pm$  0.50   &          &          &  -0.10    &  1 & \cite{Kham14}&  -7.38  \\%&  107  $\pm$  40 &   -79   $\pm$   22  &  4     $\pm$   44  &    1.1   $\pm$   0.7  &    0.86   $\pm$   0.03 \\
NGC~6791    & 0.29   & 8.3     $\pm$  0.30   &   0.010  &    0.092 &     0.124 & 32 &\cite{Linden17} &        \\%&         &                 &       &                 &    \\
NGC~6809    & -1.93  & 13.00   $\pm$  0.25   &   0.366  &    0.323 &     0.470 &14 &\cite{EugenioUVES}&   -7.55  \\%& -188  $\pm$    07 &   -202  $\pm$   28  &   -105  $\pm$   14  &    3.7   $\pm$   0.3  &    0.51   $\pm$   0.04 \\
NGC~6838    & -0.82  & 11.00   $\pm$  0.38   &   0.344  &    0.398 &     0.490 & 12 & \cite{EugenioUVES}&  -5.60  \\%&  -77  $\pm$  14 &   -58   $\pm$   10  &   -2    $\pm$   14  &    0.3   $\pm$   0.0  &    0.17   $\pm$   0.01  \\
\\%NGC~6934    & -1.54  &                  &          &          &           &    &    &   &   76  $\pm$  60 &   -521  $\pm$  48  &   -109  $\pm$   71  &    21.2  $\pm$   9.5  &    0.72   $\pm$   0.07  \\
NGC~6981    & -1.48  & 11.50   $\pm$  0.25   &          &          &        &   &  &  -7.04  \\%&          &       &                           &       &     \\
NGC~7006    & -1.46  & 11.25   $\pm$  0.25   &          &  0.350   &           & 105 &\cite{Kirby08}  &  -7.68  \\%&       &                         &          &       &            \\
NGC~7078    & -2.33  & 12.75   $\pm$  0.25   &   0.290  &    0.303 &     0.450 & 13 &\cite{EugenioUVES} &  -9.17  \\%&  -148 $\pm$    28 &   -219  $\pm$   14  &   -58   $\pm$   24  &    4.9   $\pm$   0.8  &    0.32   $\pm$   0.05 \\
NGC~7089    & -1.66  & 11.75   $\pm$  0.25   &   0.135  &    0.185 &     0.410 &94 &\cite{Bulbo17} &  -9.02  \\%&  96  $\pm$     41 &   -206  $\pm$   38  &   -320  $\pm$   49  &    4.4   $\pm$   0.3  &    0.39   $\pm$   0.06 \\
NGC~7099    & -2.33  & 13.00   $\pm$  0.25   &   0.309  &    0.331 &     0.51  & 10&\cite{EugenioUVES} &  -7.43  \\%&      &                 &       &          &            \\
Arp~2       & -1.74  & 12.00   $\pm$  0.38   &   0.32   &    0.31  &     0.38  & 2  &\cite{Mottini08}&  -5.29  \\%&        &                 &       &          &            \\
Pal~12      & -0.81  & 9.0     $\pm$  0.38   &  -0.035  &   -0.045 &     0.08  & 4 &\cite{Cohen04} &  -4.48  \\%&          &                 &       &          &            \\
Ter~8       & -2.34  & 13.00   $\pm$  0.38   &   0.220  &    0.240 &     0.47  & 7 &\cite{Eu14a} &   -5.05  \\%&           &                 &       &          &            \\
Ter~7       & -0.45  & 7.75   $\pm$   0.50   &   0.035  &   -0.020 &    -0.11  &  5 &\cite{Luca05} &  -5.05  \\%&       &                 &       &         &             \\
IC~4499     & -1.62  & 11.25  $\pm$   0.25   &          &          &       &    &  &  -7.33  \\%&          &       &                           &       &     \\
Rupr~106    & -1.78  & 10.75  $\pm$   0.25   &  -0.020  &   0.000  &     -0.02 & 9 &\cite{Sandro13} &   -6.35  \\%&        &                 &       &          &           \\
Pyxis       & -1.20  & 10.50 $\pm$    0.25   &          &          &     &   &   &   -5.75  \\%&  &                &         &                 &       \\
\hline
\end{tabular}
}
\end{table*}

%--------------------------------------------------------------------
\section{Age-metallicity relation}
Figure~1 shows the age-metallicity relation found by \cite{Leaman13}, colour-coded following a
three-group classification: old metal-poor clusters (black), clusters populating the metal-rich branch of
the AMR split (red), and clusters populating the metal-poor branch of the AMR (blue). 
This three-groups classification is based on the following considerations: i) isolate the metal-rich
branch including its oldest clusters; ii) identify the metal-poor branch clusters that are younger than the spread
of the age plateau (ranging from 13 to 12 Gyr), and iii) define a separate class for the oldest metal-poor
clusters in the AMR plateau. This arbitrary classification is established for analysis purposes. It allows me to generally compare the
properties of the clusters in the different features of the AMR \citep[but see also][for a physically motivated separation by 
accreted galaxy mass in this parameter space]{Kruijssen18}. In addition, as shown in the following sections,
the conclusions of the analysis are not influenced by slight modifications of the three groups.

As noted in the Introduction, the metal-rich (red) group is generally considered to be formed {\it \textup{in situ}}
and to be associated with the galactic disc. The young metal-poor (blue) group and at least a fraction of 
the old metal-poor (black) group are assumed to have an accretion origin. This classification into three groups is used in
the following sections to compare the chemodynamical and structural parameters of the clusters
in these groups and with respect to the local disc field stars.

\smallskip
First of all, Figure~2 again shows the \cite{Leaman13} AMR (black points) together with the 
local thick disc (light blue points) and the local thin disc AMRs (green points) from the AMBRE sample. For comparison purposes,
the oldest age limit of the field sample  was corrected to be in agreement with the GC 
one (using a constant shift of 1.5~Gyr that can be explained by modelling-dependent age biases).

On one hand, Fig.~2 shows that the metal-rich branch of the GC AMR bifurcation partially overlaps 
the locus of thick discs. This suggests that these GCs might have formed in situ during the first epochs of disc formation, in agreement with the \cite{Leaman13} 
suggestion of a disc classification of those objects. On the other hand,
the metal-poor branch clusters are, as expected, much more metal poor than the thin-disc clusters, with 
only two clusters (Pal~12 and Ter~7) overlapping the locus of the thin disc AMR.  

To better understand if a link can be established between the two GC branches and the disc bimodality, 
 the homogeneous sample of 67 halo stars investigated by \cite{Schuster12} 
was included in the plot as orange points. After \cite{NissenSchuster2010} revealed
the existence of low-$\alpha$ halo stars, a series of papers of the same group have explored
their physical properties in detail, comparing them to canonical high-$\alpha$ halo stars (see also the recent
work of \cite{Hayes18}. This confirmed the existence of two chemically distinct halo populations in the field
halo stars. In particular, \cite{Schuster12}
investigated the mean ages of a sample of high-$\alpha$ and low-$\alpha$ stars as a function of the metallicity (cf. their Table~2).
Figure~2 presents these mean values for the high-$\alpha$ stars as orange squares and for the low-$\alpha$  ones as orange  circles (the same shift of 1.5~Gyr as was applied to the disc field sample was used).
The error bar corresponds to the reported standard deviation at each metallicity bin.
The \cite{Schuster12} stars again present a bimodality. Low-$\alpha$ halo objects seem to approximately lie along 
the metal-poor GC branch and certainly below the metal-rich branch and the thick-disc locus, suggesting a connection
with the thin-disc locus at higher metallicities. Nevertheless, the lack of objects in the younger
part of the AMR bimodality between -1.2 and -0.8~dex prevents me from robustly concluding about
the connection between the metal-poor branch GCs and thin-disc GCs.  
The high-$\alpha$ halo stars, as already known, more or less
overlap the thick-disc sequence and the GC metal-rich sequence.

Generally speaking, the GC AMR bimodality seems to have some overlap with the Galactic
disc AMR bimodality, especially during the thick-disc formation phase. 
However, the inhomogeneity of the plot, which includes three different sets of data (clusters, disc, and field halo stars), 
each of it with its biases in age and metallicity, prevents a conclusion on the relations between them.  Despite this,
Fig.~2 shows that a bimodality of the AMR is present in the three galactic populations.
A more homogeneous
data set that includes clusters, field halo stars, and disc stars needs to be analysed to clearly conclude 
on the substructure of the galactic AMR.

Finally, it is also important to point out that substructures in the age-metallicity plane, such as the above discussed bimodalities, 
can be created by different evolutionary processes that differently different structural galactic
components independently. For instance, accretion from dwarf galaxies in the halo or the disc, gas infall or radial outflows in the disc,  radially dependent star formation
rates, etc. might all be responsible for an AMR split in a way that the overlap of the clusters and disc features would only be
result of a degeneracy in the age-metallicity domain and not the expression of a common evolutionary path. 
Despite this caveat, the comparative analysis of different galactic components is extremely useful to allow
a general vision of the Milky Way evolution and of the complex relations between its different stellar populations.

\begin{figure}[ht]
\includegraphics[width=9cm,height=6cm]{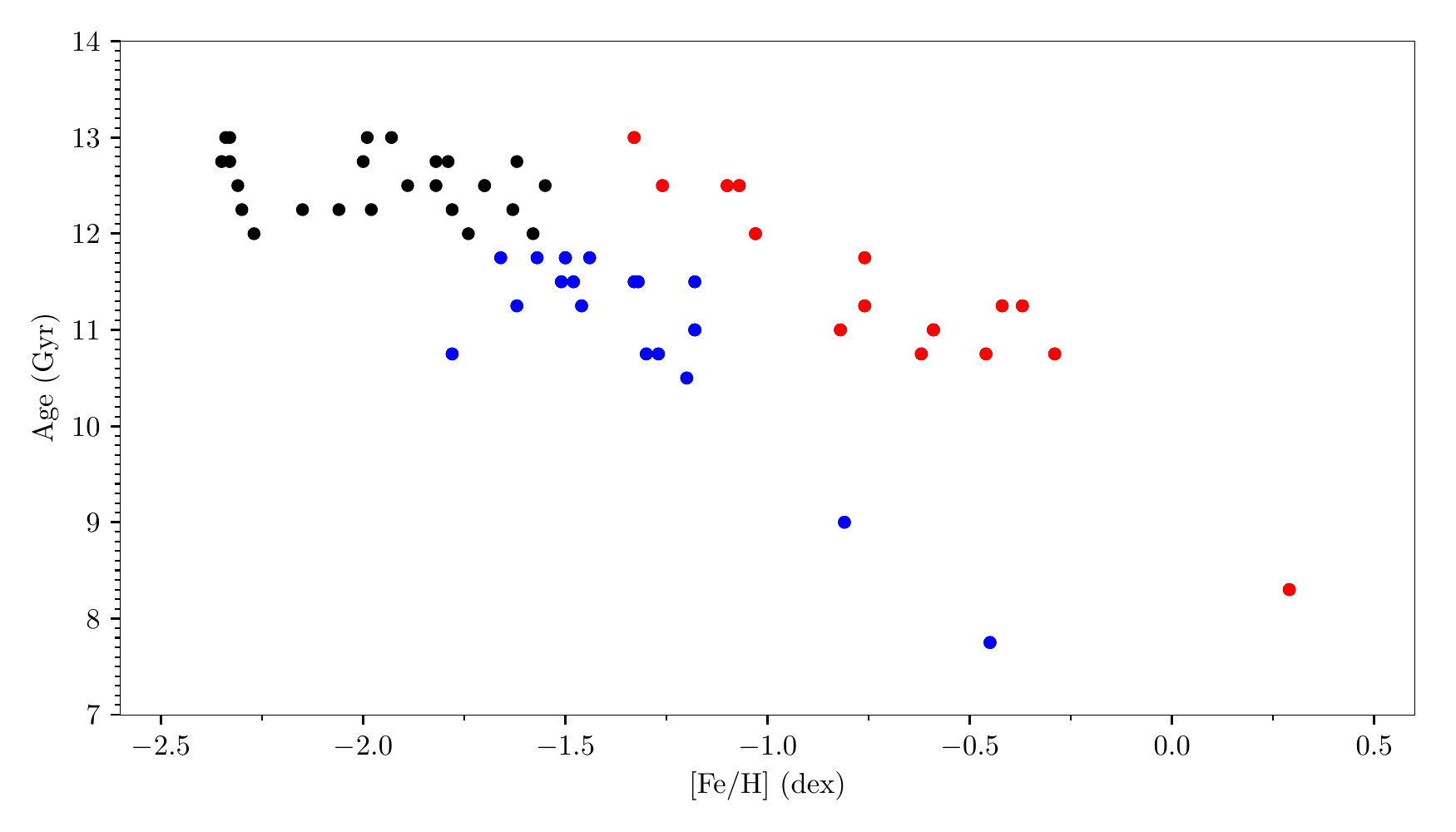}
\caption{\cite{Leaman13} age-metallicity relation, colour-coded to define three groups of clusters:
old metal-poor clusters (black), metal-rich branch clusters (red), and metal-poor branch (blue) clusters.}
\label{Groups}
\end{figure}
% Produced with the macro: /home/arecio/DiscGCs/AgesMet2branches.py

\begin{figure}[ht]
\includegraphics[width=9cm,height=7cm]{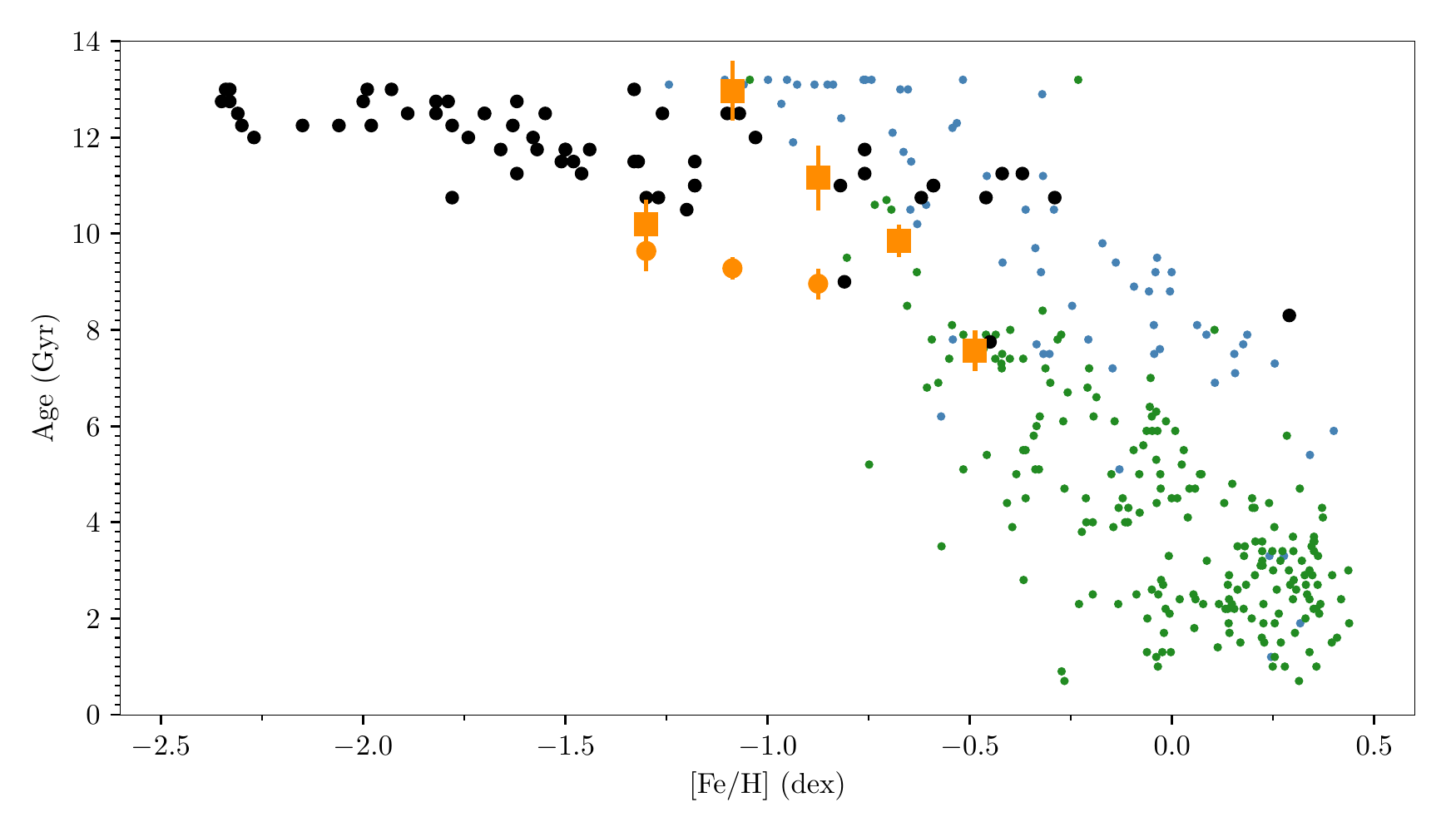}
\caption{\cite{Leaman13} AMR (black points) together with the 
local thick disc AMR (light blue points) and the local thin disc AMR (green points) from \cite{MichaelAmbre} and the mean values for
the low-$\alpha$ (orange circles) and high-$\alpha$ (orange squares) halo field stars investigated by \cite{Schuster12} (orange).} 
\label{AgeMetDisc}
\end{figure}
% Produced with the macro: /home/arecio/DiscGCs/AgeMet.py

%---------------------------------------------
\section{$\alpha-$element abundances and the GC AMR bifurcation}
In this section the mean $\alpha$-element abundance of GCs is examined
in relation with the AMR bifurcation. It is well known that the \aabun~ chemical abundance ratio is an important indicator of
the chemical evolution of a system. In particular, and according to the time-delay model \citep[][]{Tinsley79, MatteucciGreggio},
the initially enhanced $\alpha$-abundance levels with respect to iron start to strongly decline with \feh~ after the supernovae 
Ia explosion rate reaches maximum. 

On one hand, this produces a knee in the \aabun~ versus \feh~ trend 
whose location provides constraints on the star formation rate during the early star formation
of a system \citep[e.g.][]{deBoer14}. As a consequence, the knee location also depends on the system total mass: the less massive the system, the lower the \feh~ value of the \aabun~ turnover. Observations of dwarf galaxies of
different masses have confirmed this dependence. A low-mass galaxy like Carina shows a very metal-poor knee (\feh$=-2.7 \pm 0.3$~dex), while the more massive Sculptor presents a slightly more metal-rich knee \citep[\feh$=-1.9 \pm 0.1$~dex,][]{McConnachie12}. Finally, the knee of higher mass galaxies, like Sagittarius can reach metallicities as high as \feh$=-1.27 \pm 0.05$~dex \citep[][]{deBoer14}. In addition, the Milky Way galaxy has a knee in the range -1.0~dex to -0.5~dex, depending on the authors
\citep[][]{deBoer14, Alvaro2}

On the other hand, the initial mass function (IMF) of the system influences the primordial \aabun~ and
therefore the \aabun~ abundance of the low-metallicity plateau.
Finally, the \aabun~ is a fairly good age indicator for \feh~ values higher than the knee value \citep[e.g.][]{Haywood2013, MichaelAmbre}.

Figure 3 presents \mgfeabun~ as a function of \feh~ for i) the \cite{Leaman13} GCs with 
available chemical information, using the colour code of Fig.~1, ii) the AMBRE thick- and thin-disc comparison
sample, and iii) the low-$\alpha$ and high-$\alpha$ halo field stars investigated by \cite{Schuster12}. For clarity, the same colour code as in Fig.~2
is used. Clearly, the GCs show a bimodal behaviour 
for metallicities higher than about -1.5~dex, although a large scatter is present. Moreover, 
this bifurcation separates the two cluster populations that were identified in the AMR split: clusters in
the metal-rich branch (red points in Figs.~1 and~3) seem to have higher \mgfeabun~ values
than clusters in the metal-poor branch (blue points in Figs.~1 and~3). This would be
in agreement with a possible accreted origin of the metal-poor young clusters, as suggested by
\cite{Leaman13}. In addition, when the two disc sequences are considered, the thick-disc sequence presents \mgfeabun~ abundances that are compatible with those of the metal-rich branch clusters, again in agreement
with a disc origin for these clusters. The thin disc, for its part, shows higher \mgfeabun~ abundances than the
youngest GCs in the metal-poor branch, although both sequences seem to join at
around \feh$\sim$-1.0~dex. Finally, the halo field low-$\alpha$ stars occupy the same locus as
the more metal-rich clusters of the metal-poor branch, while the \mgfeabun~abundances of high-$\alpha$ halo field stars are
more similar to those of the thick disc and the metal-rich branch clusters (declining for
metallicities higher than about -0.75~dex).
As a consequence, the three studied populations (GCs, disc, and halo field stars) 
and their sub-classes (metal-rich and metal-poor branch clusters, thick and thin disc, and high- and low-$\alpha$ halo stars)
show similar overlaps in the AMR and in the \mgfeabun~ versus \feh ~ planes.
It has to be stressed, nevertheless, that the disc, the GC, and the halo 
sequences are plotted together mainly for a general comparison, as data biases between the two data sets
prevent a robust conclusion on their links. In addition, the origin of the \cite{Schuster12} halo stars is under debate, so that 
their abundance patterns alone do not provide evidence whether the halo and its GCs formed in situ
or ex situ.

\begin{figure}[ht]
\includegraphics[width=9cm,height=6.5cm]{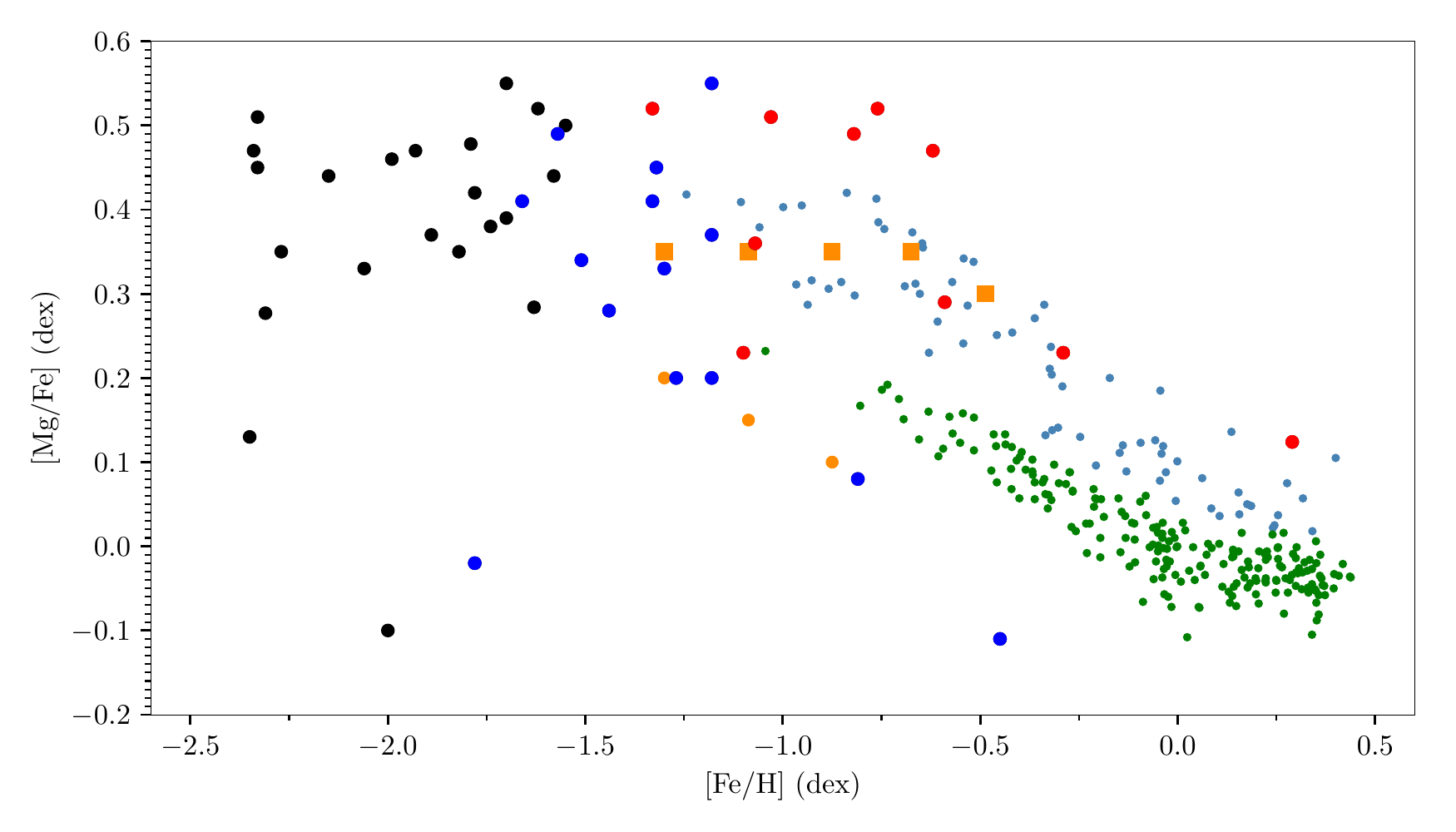}
\caption{\mgfeabun~ abundance as a function of \feh~ for the  \cite{Leaman13} GCs, the
disc comparison sample, and the  \cite{Schuster12} halo field stars. The colour codes are those of Figs.~1 and~2  for the clusters, the disc, and the halo
field stars.}
\label{MgFeTous}
\end{figure}
% Produced with the macro: /home/arecio/DiscGCs/MgFe.py

Furthermore, only the mean chemical abundances for each cluster are taken into account in Fig.~3,
neglecting the well-known internal light element abundance scatter.   GC star formation is currently believed to be bimodal,
with a second generation of stars born from the ejecta of the primeval generation \citep[e.g.][]{EugenioUVES}. 
There is evidence that the Mg-Al cycle is active in cluster polluters, causing the so-called star-to-star anti-correlation between the Mg and the Al abundances.
In this sense, part of the scatter observed in the GC data in Fig.~3 could come from this internal spread in the
Mg abundances. To solve this problem, Fig.~4 shows the \aabun~ versus \feh~ of the \cite{Leaman13} GCs,
considering four different $\alpha$-elements (Mg, Si, Ca, and Ti). The points are colour-coded by their age. Clearly,
the scatter in the more metal-poor clusters is reduced with respect to Fig.~3. In addition, the bifurcation of the abundances
at about \feh$\sim$-1.5~dex, although still present, seems to be less clear, with several metal-poor branch clusters having
high \aabun, compatible with those of the older metal-poor population.

\begin{figure}[ht]
\includegraphics[width=9cm,height=6.5cm]{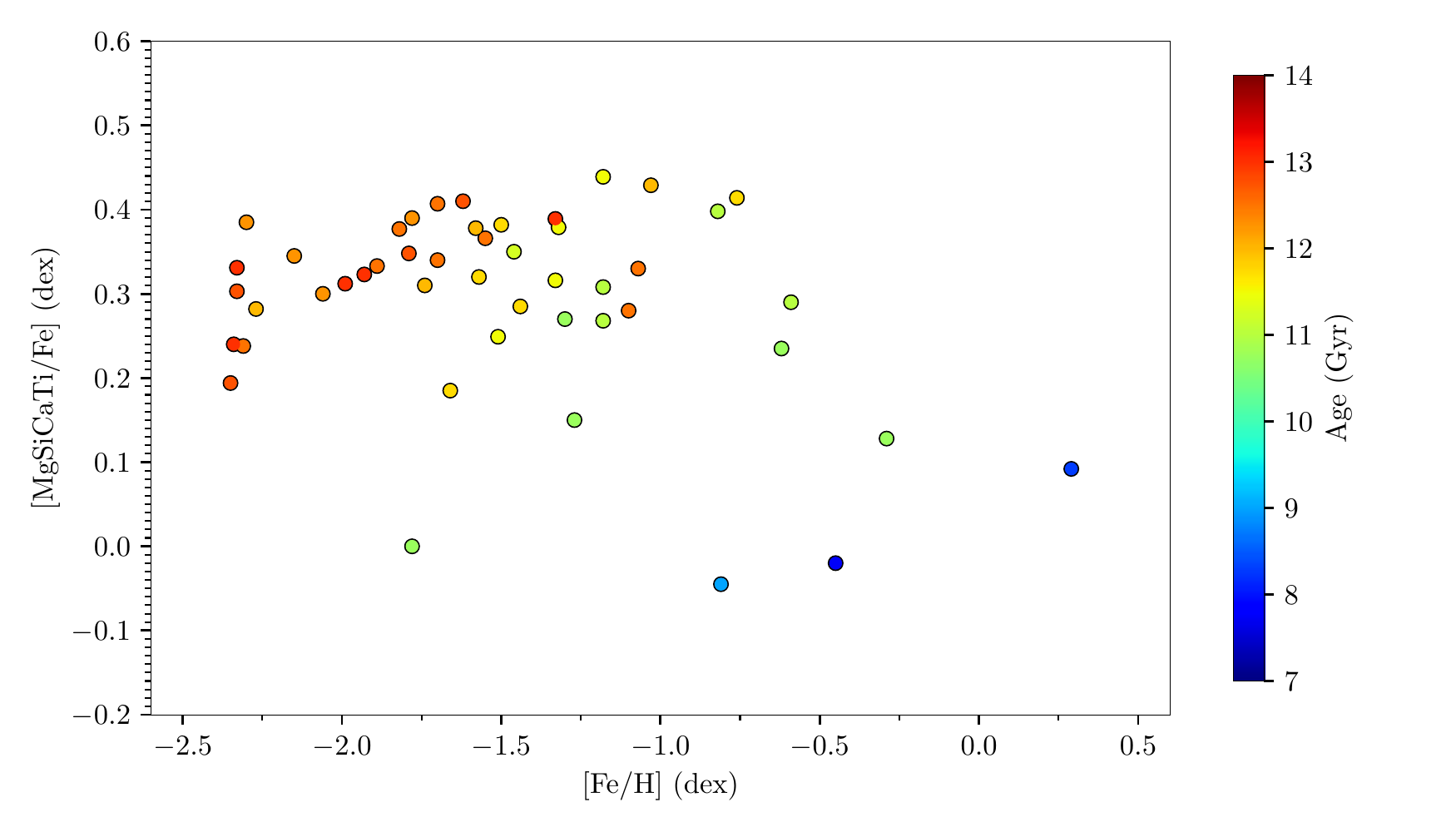}
\caption{Mean \aabun~ abundance vs. metallicity for the \cite{Leaman13} GCs,
considering four different $\alpha$-elements (Mg, Si, Ca, and Ti). The points are colour-coded by cluster age.}
\end{figure}

Finally, Fig.~5 shows the mean \aabun~ abundances considering only Si and Ca. On one hand, this allows avoiding 
the scatter due to the 
Mg-Al anti-correlation, and on the other hand, it excludes Ti, which is not a pure $\alpha$-element and is also produced  in great abundance
by Type~Ia supernovae. The Si and Ca mean abundance clearly presents a less strongly scattered high-$\alpha$ sequence of
clusters from -2.5~dex to -1.0~dex. The mean value of the \sicafeabun~ of this sequence is 0.35~dex, with only 0.05~dex
of standard deviation. This observational scatter is perfectly within the errors of the abundance estimate, especially for an inhomogeneous sample taken 
from the literature. As a way of comparison, the dispersion for the same clusters in the range -2.5~dex $\le \feh \le$ -1.0~dex is 0.11~dex in \mgfeabun~ and \aabun.
Moreover, Fig.~5 shows a knee around $\sim$-0.75~dex, from which a declining sequence of abundances starts that is
similar to the thick-disc sequence. Finally, only a few metal-poor clusters present low \sicafeabun~ abundances (Rupr~106, NGC~7089, NGC~1261, and Pal~12).

\begin{figure}[ht]
\includegraphics[width=9cm,height=6.5cm]{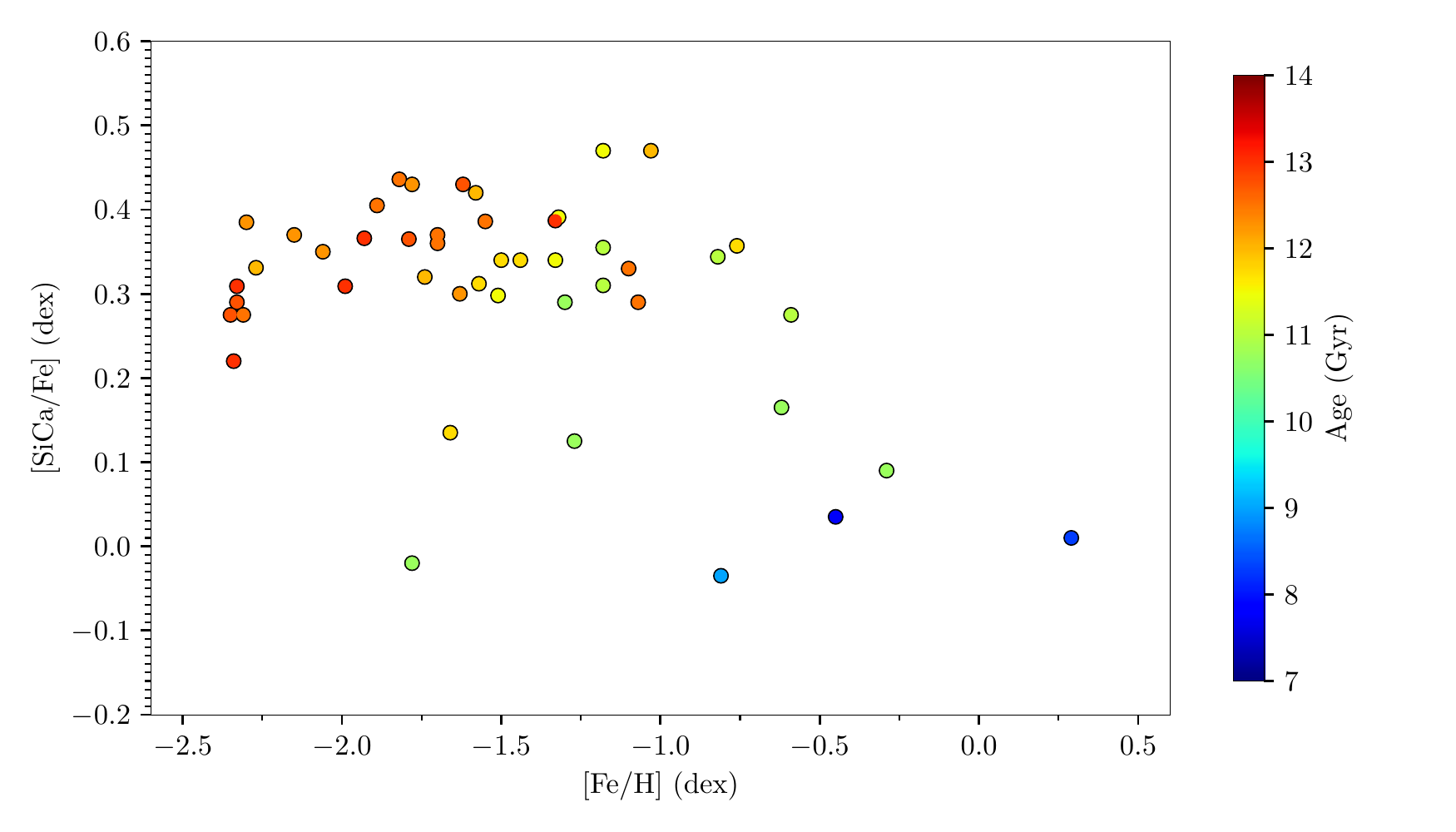}
\caption{Mean \aabun~ abundances as a function of metallicity for Si and Ca alone. The age colour-code is the same
as in Fig.~4.}
\end{figure}
% Produced with the macro: /home/arecio/DiscGCs/SiCaClusters.py
 
 \begin{figure}[ht]
\includegraphics[width=9cm,height=6.5cm]{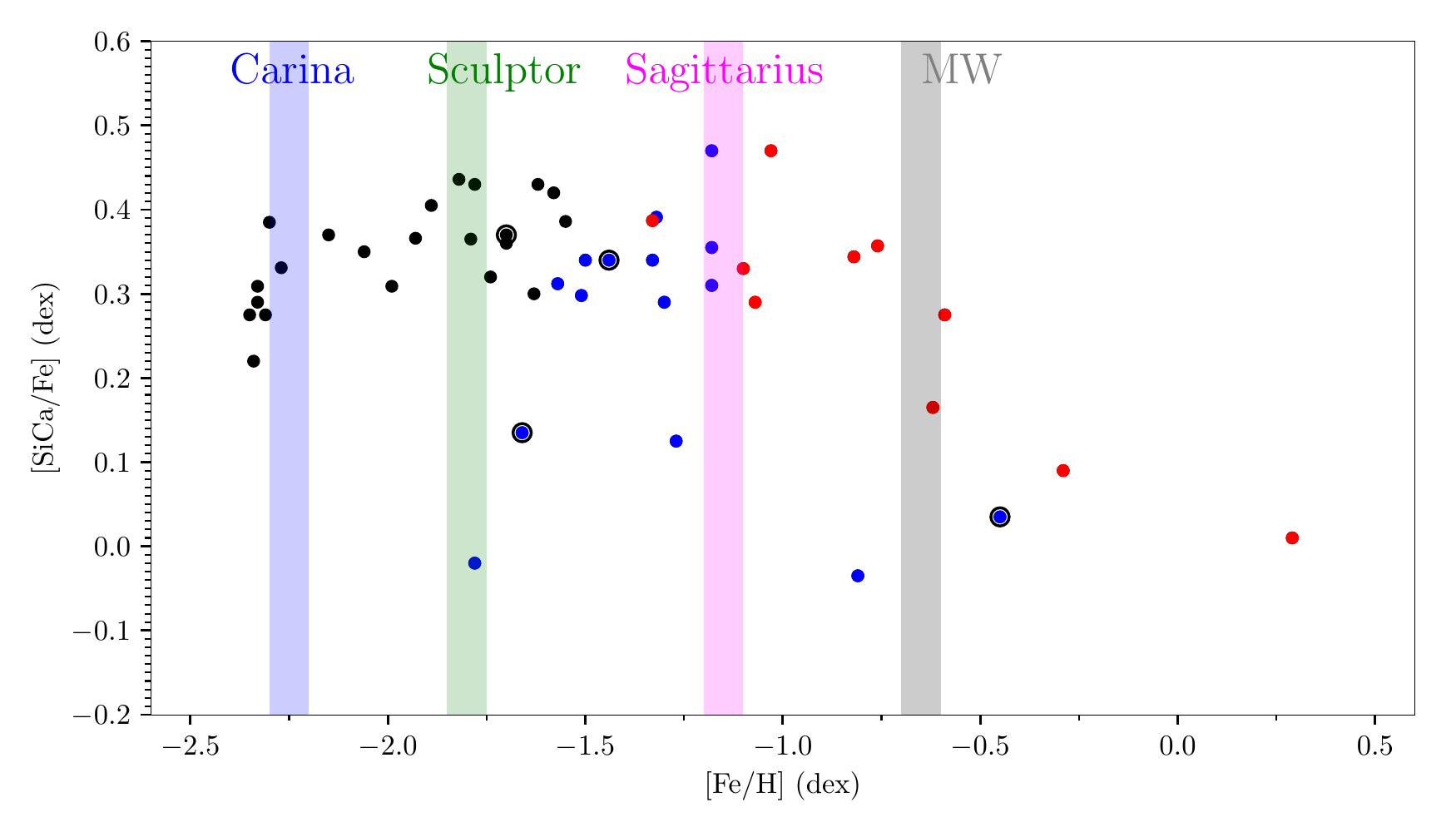}
\caption{Same as Fig.~5, but using the colour code of Fig.~1 to distinguish the different groups of clusters in
the AMR. The knees of the Milky Way, Sagittarius, Sculptor, and Carina are presented as vertical bands
in grey, pink, green, and blue, respectively.}
\end{figure} 
% Produced with the macro: /home/arecio/DiscGCs/SiCaMP.py
 
 To facilitate comparison between the \sicafeabun~ abundance distribution and the AMR cluster groups, Fig.~6 
shows the same colour-coding as Fig.~1. Interestingly, the majority of the metal-poor branch clusters present high
values of the \sicafeabun~ abundance, equal to (within the error estimates) those of the oldest in situ population
and those of the co-eval clusters in the metal-rich branch. To facilitate interpretation of this result, Fig.~6 includes
the knee positions of the Milky Way (grey vertical band) and those of three different dwarf galaxies following \cite{deBoer14}
and \cite{Tolstoy09}: Sagittarius, Sculptor, and Carina.

%The clusters in this high-$\alpha$ sequence do not seem to be formed in low star formation
%rate environments. In addition, 
Several conclusions can be inferred from Figures~5 and 6. 
First of all, as I described above, the
metal-rich branch GCs (in red) follows the Milky Way thick-disc sequence, with a knee at about \feh$\sim$-1.0~dex.
Secondly, the majority of the metal-poor clusters have  high \sicafeabun~ abundances with a very low scatter. This indicates
that if they have been formed in dwarf galaxies and were subsequently accreted by the Milky Way, two possibilities exist:
1) they could all have been formed in massive dwarf galaxies like Sagittarius, which presents a knee in the intermediate-metallicity regime
(\feh~ at about -1.2~dex), or 2) they could have been formed in dwarf galaxies with a variety of masses, but have been accreted very early (especially
the low-mass dwarfs), before the onset of Type~I supernovae causes the decrease in \sicafeabun~ ratio. Conversely, 
there is also room for an {\it \textup{in situ}} formation of all the metal-poor clusters in the Milky Way halo, which has a metal-rich \sicafeabun~ knee.

Finally, four GCs that are known to be non-monometallic, that is, to have an iron abundance spread \citep[][]{Grebel16} (NGC~6656, NGC~6715, NGC~7098, and Terzan~7),
are labelled in Fig.~6 with an open black circle. Two of them, Terzan~7 and NGC~7089, clearly present low \sicafeabun~
abundances, which is an unmistakable sign of accretion. In addition, the cluster NGC~6715 (M54), which is known to be embedded at the centre of the Sagittarius dwarf galaxy, presents a high \sicafeabun~ ratio that is compatible with the high-metallicity knee of the massive
Sagittarius.

In conclusion, when $\alpha$-element abundances (not disturbed by the internal scatter of GCs) are considered,
the hypothesis of an accretion origin for the metal-poor branch clusters seems to be restricted, with only a few exceptions, to the most massive dwarf galaxies and to early accretion of low-mass dwarfs. On the other hand, and based on the $\alpha$-element abundances alone,  their in situ\textup{} formation
cannot be ruled out. Their  $\alpha$-abundance patterns are indeed also compatible with the isolated chemical evolution of a
high-mass galaxy such as the Milky Way.

\section{Total luminosity distributions}

 This section analyses the total visible luminosity of the clusters, taken from the most recent version of the Harris catalogue \citep[][]{Harris}. The luminosities are
compared with the different groups defined in Sect.~3 from the AMR.
This exercise is nevertheless challenging because of the complex physics involved in the formation and disruption of
GCs \citep[e.g.][]{Forbes18}, which hinders the interpretation of the cluster luminosity function.
On one hand, the luminosity function of GCs has historically been considered universal and was used as a secondary distance indicator.
Recent results have revealed weak dependences on Hubble type, mass, environment, and dynamical history 
of the host galaxy \citep[e.g.][]{Marina12, Harris14}, which affect the turnover luminosity (where the luminosity 
function peaks), only to second order.   On the other hand, the luminosity function of Local Group dwarf spheroidals and that of the outer halo Milky Way cluster population
are found to contain  fainter GCs \citep[][]{vdB06, VdBMakey04, Eu10PCA}.

%The total luminosity of a globular cluster can be used as a good proxy of its total mass.
With these challenges in mind, the luminosity distribution of the clusters in the \cite{Leaman13} sample can still be
studied to report similarities or differences depending on age and or metallicity. It is important to note that the great majority of bulge clusters are not included in the \cite{Leaman13} data base. This biases the analysis of the metal-rich population and excludes very high density environments. 

First of all, the top panel of Fig.~7 shows the (absorption-corrected) M$_{V}$  for the old GCs in the AMR plateau{\it } (grey histogram) compared to that
of the metal-poor (blue histograms)  and metal-rich branch clusters (shown in red).  As it is difficult to
infer the corresponding mass loss of each GC that is lost in the stars that have escaped the cluster, the three histograms
in the top panel were smoothed through a kernel density estimation (KDE) technique. To this purpose, two different
$\lambda$ covariance factors were used: 0.2 (middle panel of Fig.~7) and 0.5 (lower panel). Finally, a Kolmogorov-Smirnov (K-S)
test was performed to compare the three distributions. When the M$_{V}$ distribution of the metal-poor
branch young GCs (shown in blue) is compared that of the metal-poor old GCs (shown in grey), the K-S p-value is 0.65, showing that both
distributions are very similar. This result highlights the fact that no particular discontinuity in the typical  luminosity of the clusters seems to exist 
among the metal-poor clusters, even considering an age range of at least 3 Gyr.
Conversely, metal-rich branch GCs (in red) are generally  less luminous than the others: the total magnitudes of 80\% of the metal-rich branch clusters are fainter than -8.0, while this number decreases to 50\% and 55\% for the other two
subpopulations I analysed (metal-poor branch and old clusters).
The K-S p-value that compares the two distributions is 0.33, which quantifies the lower degree of similarity between
the disc-associated clusters and the old halo discs.
Nevertheless, a final word of caution has to be said about two aspects of this analysis:
i) there is not necessarily a mass dependence if a difference in Mv is seen because of the difficult interplay
between cluster mass and luminosity, which depends on multiple parameters (metallicity, environmental effects, etc.), and
ii) the reported K-S p-values are estimated in a regime of relatively low-number statistics.

\begin{figure}[ht]
\includegraphics[width=9cm,height=8cm]{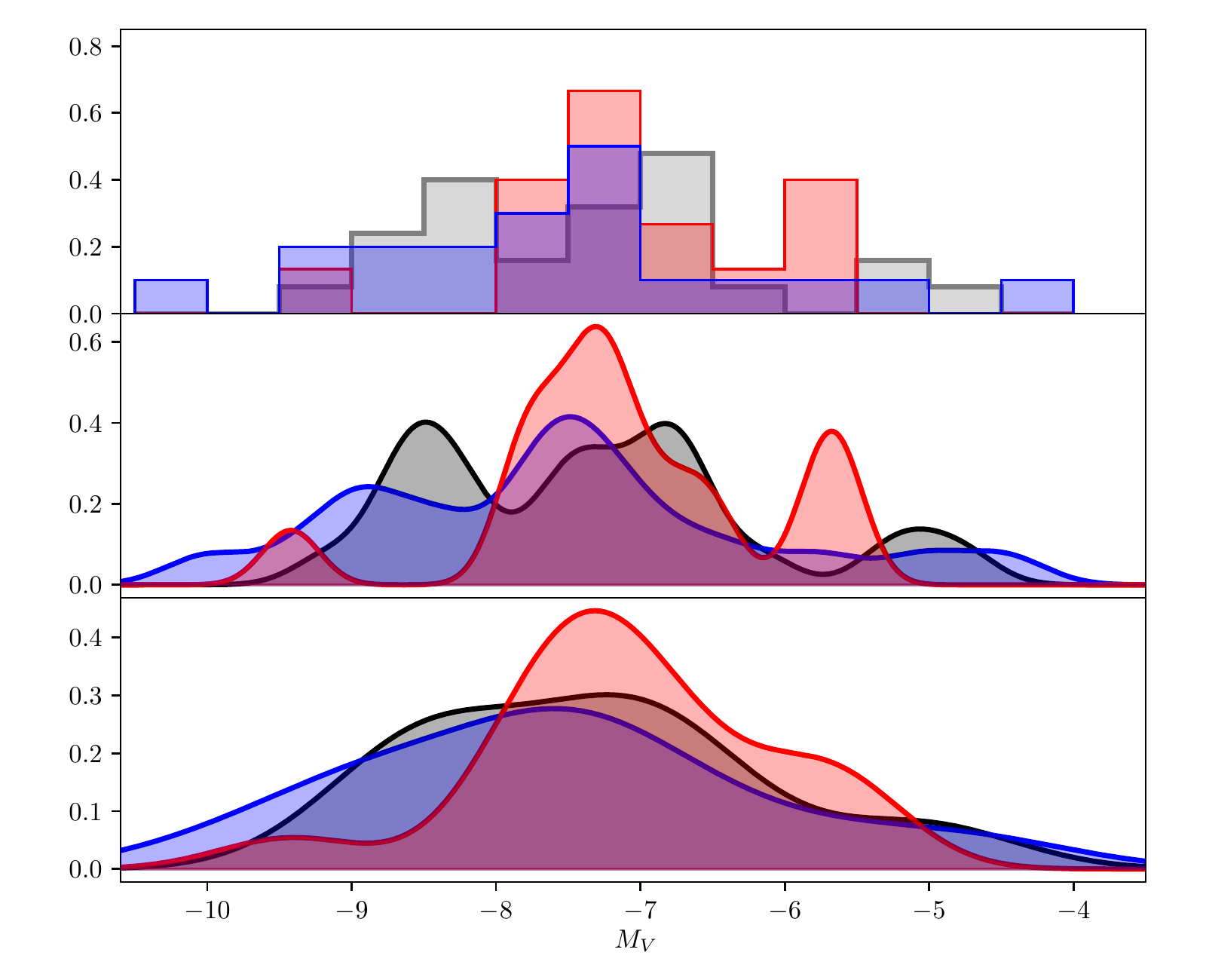}
\caption{Top panel: Normalised M$_{V}$  distribution of old in situ GCs (grey histogram) compared to that
of the metal-poor branch clusters (blue histogram) and that of the metal-rich branch clusters (shown in red). Middle panel and
lower panels: Three distributions smoothed through a KDE, with a $\lambda$ covariance factor of 0.2 and 0.5.}
\end{figure} 
% Produced with the macro: /home/arecio/DiscGCs/MvtHisto.py
 
 \smallskip
 To further examine the nature of the possible luminosity distribution differences between the clusters subpopulations, Figure 8
 explores the chemical dependences on cluster luminosity.
 The upper panel shows the mean \sicafeabun~ as a function of the total cluster magnitude, using a colour code with age.
 On one hand, as I highlighted in the previous section, the majority of the clusters appear in a high-$\alpha$ regime (\sicafeabun$\sim$0.35~dex).
 No particular tendency with Mv or age appears for these clusters, which offers no additional constraints about their origin (e.g. massive dwarf galaxies, early accretion 
 of low-mass dwarfs or in situ{\it } formation).
On the other hand, a less densely populated low-$\alpha$ regime (\sicafeabun$\le$0.20~dex, with a mixture of metal-rich and metal-poor clusters) is also visible and 
appears to be clearly separated from the standard high-$\alpha$ regime, especially for the brighter luminosities. %Those low-$\alpha$ clusters present a dependency of \sicafeabun~ with Mv (r$=$-0.7), with more luminous
 %clusters having slightly higher \sicafeabun~ values. In addition, they present a strong age dependence with Mv (r$=$-0.85),
 %with more luminous clusters being older. No correlation with metallicity seems to exist.
 
  To distinguish the nature of the features seen in the \sicafeabun~ versus Mv plane, the lower panel of Fig.~8 presents the 
 mean \sicafeabun~ as a function of the total cluster magnitude with a colour code showing clusters in the AMR metal-poor branch (blue), the
 metal-rich branch (red) and the old metal-poor population (grey). 
First of all,  {\it \textup{high}-$\alpha$} metal-poor branch clusters (which represent the majority of this population and cover the
 metallicity regime between about -1.6 dex and -1.1~dex) are all brighter than
 about Mv$\sim$~-7~mag. This is not the case for more metal-poor old clusters (\feh $\lessapprox$ -1.5~dex, grey points), which can be as faint as -4.5~mag.
 As discussed in the previous section, the high-$\alpha$ metal-poor population could have formed in a mixture of massive and less massive dwarf galaxies,
 in contrast to the high-$\alpha$ population at normal metallicity level, whose external origin requires higher-mass satellites. 
 The greater luminosity range spanned by the old metal-poor clusters with respect to the high-$\alpha$  clusters  in the intermediate metallicity range suggests 
 a higher heterogeneity of formation (and eventually, distruption) environments at lower metallicity.
  
  Secondly, when the few clusters with \sicafeabun$\le$0.20~dex are also considered, 
  metal-poor branch clusters present an anti-correlation (r$=$-0.59, solid blue line) of \sicafeabun~ with Mv, while this is not the case for metal-rich branch clusters (r$=$-0.08, solid red line). 
 This difference between the two subpopulations is strongly driven by the fact that no faint high-$\alpha$ clusters are observed among 
 the metal-poor branch subpopulation, in contrast to the metal-rich branch population.
 Although the reasons for such a difference can be multiple and degenerate, and the number of clusters with available data is not very
 high, I point out that clusters that are separated in the AMR plane seem different in their chemical properties as well when the low-luminosity
 regime is considered. The environment in which faint metal-poor branch  clusters have formed and evolved 
 could have been subject to lower star formation rate or preferential loss of elements from massive supernova type II  than the environment that hosts
 faint metal-rich branch clusters. These two conditions are compatible with the formation of the faint metal-poor branch clusters in
 dwarf galaxies, and therefore are compatible also with their accreted origin. In addition, the observed anti-correlation of the \sicafeabun~ ratio with Mv for
 metal-poor branch clusters is also puzzling, suggesting that their chemical evolution conditions might be different for brighter clusters 
 (higher star formation rate or more efficient massive supernova feedback) than for fainter clusters. Although the available data
 do not allow concluding on this, one possibility would be that more massive dwarfs would host more luminous clusters, as has been suggested
 by \cite{vdB06}.

\begin{figure}[ht]
\includegraphics[width=9cm,height=10cm]{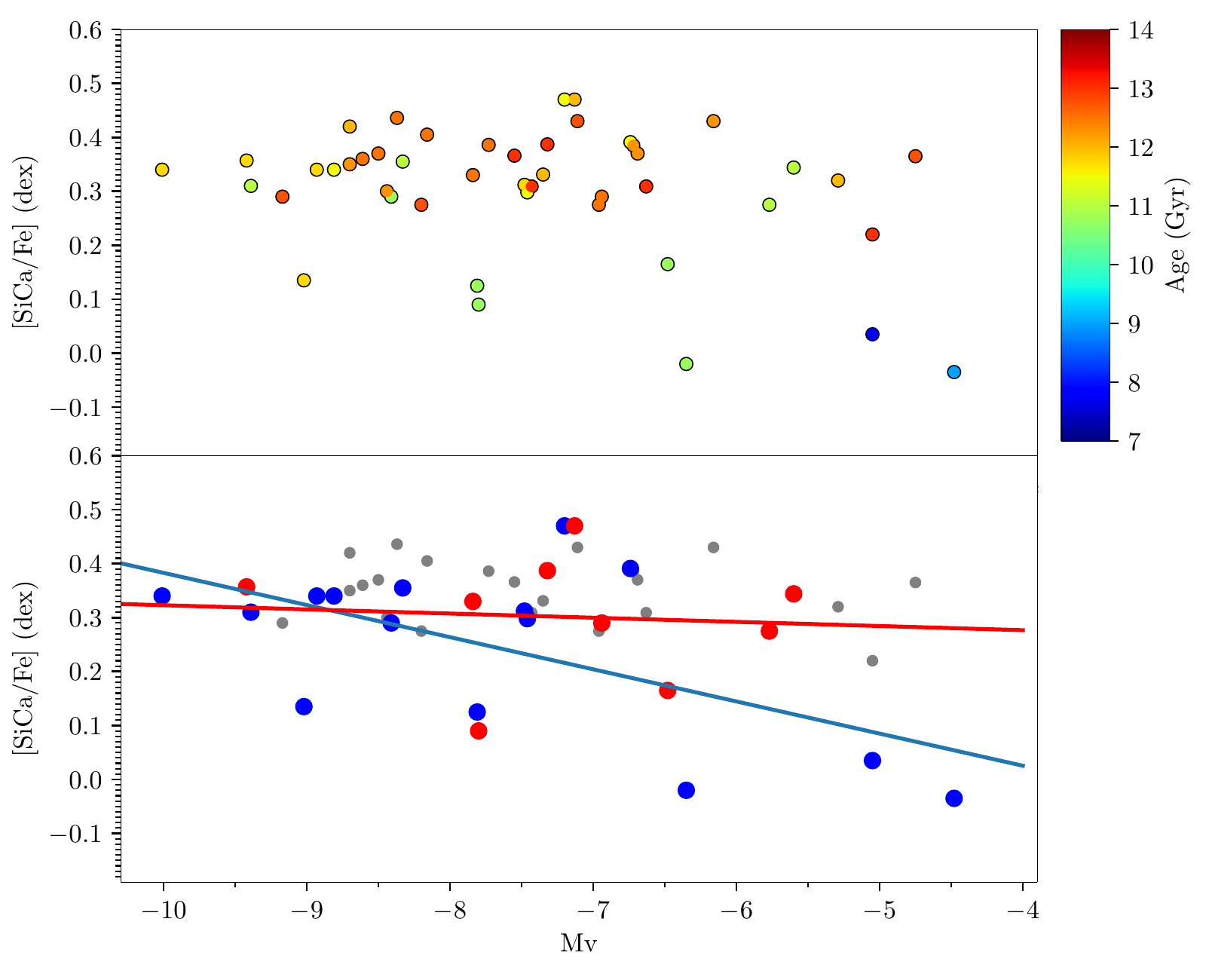}
\caption{Top panel:  Mean \aabun~ abundances as a function of total cluster magnitude for Si and Ca alone. Points are
colour coded by age. Lower panel: Same as the upper panel, but identifying clusters in the AMR metal-poor branch (blue points),
the AMR metal-rich branch (red points), and the old metal-poor population (black points). The solid blue and red lines
correspond to the linear fit of the metal-poor and the metal-rich branch clusters, respectively. 
An anti-correlation of  the \sicafeabun~ abundance with the total magnitude (r$=$-0.59) seems to be present only for
metal-poor branch clusters. }
\end{figure}  
 % Produced with the macro: /home/arecio/DiscGCs/SiCaMv.py
 
\section{Summary and discussion}

Different aspects of the galactic GC populuation are revealed by the analysis of the substructures in the Milky Way GC age-metallicity relation, coupled
with chemical abundance information, the study of the total luminosity distribution, and the comparison with
field stars data.
First of all, the bimodality in the age-metallicity relation, reported by \cite{Leaman13}, seems to overlap in
time with the halo field star bimodality \citep[][]{Schuster12} and the thick-thin disc bimodality \citep[][]{MichaelAmbre},
although the biases between the different datasets prevent us from a detailed conclusion on their relative links.

Secondly, when the $\alpha$-element abundances, which are less strongly affected by internal light-element spread of
GCs, are considered, a very low observational scatter among the metal-poor clusters is observed. A plateau at \sicafeabun$\sim$0.35~dex, with a dispersion of only 0.05~dex, is observed up to a metallicity of about
-0.75~dex. Only a few metal-poor clusters in this metallicity interval (Rupr~106, NGC~7089, NGC~1261 and Pal~12) 
present low \sicafeabun~ abundances. Moreover, metal-rich globular clusters follow the Milky Way thick-disc sequence, 
with a knee at about \feh$\sim$-0.75~dex. 
This result places a clear constraint on scenarios of GC formation
in the Milky Way. On one hand, if a substantial fraction of galactic GCs has an external origin, they must have formed either 
in galaxies that were massive enough to ensure high levels of $\alpha$-element abundances even at intermediate metallicity
(e.g. the Sagittarius dwarf galaxy), or %, in the low metallicity domain, 
in lower-mass dwarf galaxies that accreted in their early formation phase (before they reached their \aabun~ vs. \feh~ knee). 
On the other hand, the in situ{\it } formation of clusters with high \aabun~ values is also a plausible
explanation.

Finally,  the study of the absolute luminosity distribution in the different features of the AMR reveals that
the total luminosity distributions of old metal-poor GCs in the AMR plateau are similar  to those of younger clusters of intermediate metallicity. However, when only high-$\alpha$ clusters are considered,
old metal-poor clusters seem to span a wider luminosity range than younger clusters of intermediate metallicity. In addition, this lack of
faint high-$\alpha$ clusters in the AMR metal-poor branch subpopulation also contradicts what is
observed for the metal-rich branch population and reinforces the differences between these two AMR branches.
Moreover, metal-rich branch clusters are generally less luminous than metal-poor clusters.

In conclusion, although the interpretation of the differently biased data samples
is complex and multi-parametric, the analysis reported here places simple observational constraints on
scenarios of GC formation and disruption:
\begin{itemize}

\item The duplicity of the Milky Way GC population, illustrated by its bifurcated age-metallicity relation, 
is confirmed by the combined analysis of the \aabun~ abundances and the total luminosity distributions. 
Some overlap with the disc population in the various bimodalities is observed, although common evolutionary paths 
for the disc and the GC populations are not guaranteed because of the observational biases and
the degeneracy in the effects of different physical evolutionary processes. 

\item The external origin of metal-poor branch clusters seems reinforced (or at least not excluded) by the \aabun~ abundances.

\item The greater luminosity range that is spanned by the old metal-poor clusters with respect to the high-$\alpha$ metal-intermediate clusters suggests 
 a higher heterogeneity of formation environments at lower metallicity, which might reflect the contribution of low-mass
 satellite accretion.
 
 \item Accretion of high-mass satellites, as a major contribution to the current Milky Way GC system in the metal-poor and intermediate-metallicity 
 regimes, is compatible with the observations.

\end{itemize}

Generally speaking,  the duality of the GC in situ{\it } versus accretion formation scenarios remains only partially unveiled, but the combined analysis
of the AMR, the chemical abundances, and the cluster luminosities gives some constraints on the accretion epochs and/or the masses
of the accreted objects. Precise dynamical data are of course another crucial piece of the puzzle. The Gaia mission, from its second data release, 
has already started to open new paths of exploration. In particular, several studies have suggested the accretion of a high-mass satellite to which several GCs are associated, which might have built up the halo inner regions and perturbed the primordial disc \citep[][]{AminaEnceladus, Kruijssen18,MyeongSausage}.
%Moreover, \cite{Hayes18} have recently confirmed the existence of two chemically distinct halo populations in the field
%halo stars, separated in the  \mgfeabun~ vs. \feh~ plain. Their result reinforces the pioneer work of \cite{NissenSchuster2010} 
%for the solar neighborhood and recalls the composite nature of the halo populations. 
Moreover, the age-metallicity relation of halo field stars needs to be studied and placed in relation with the GC AMR in order
to confirm a possible, and expected, link between field stars and clusters.

\begin{acknowledgements}
I acknowledge financial support from the ANR 14-CE33-014-01.
I thank all the people involved in the AMBRE project for the use
of the AMBRE:HARPS data as a comparison disc sample.
I am sincerely grateful to the anonymous referee of this paper
for their careful reading of the work and for all the suggestions for
improvement. In particular, their help has been crucial for the results in Sect. 5. 
I warmly thank Patrick de Laverny for his constant support, personal
and professional.

\end{acknowledgements}

\bibliographystyle{aa} % style aa.bst
\bibliography{biblio}

\end{document}